\documentclass{PoS}

\title{Possible physics scenarios behind cosmic-ray "anomalies"}

\ShortTitle{On cosmic-ray "anomalies"}

\author{\speaker{Pasquale Dario Serpico}
\\
        LAPTh, Univ. Savoie Mont Blanc, CNRS, B.P.110, Annecy-le-Vieux F-74941, France\\
        E-mail: \email{serpico@lapth.cnrs.fr}}

\abstract{Direct techniques for cosmic ray observations have reached an unprecedented level of precision, unveiling
fine-details of the energy spectra. I will introduce the evidence for new spectral features which
has been accumulated by new experiments over the past few years, and review the main ideas invoked
in the theoretical explanations of the revealed spectral breaks and elemental spectra non-universality.
I will also briefly comment on the complementary situation of antimatter observations.
}

\FullConference{The 34th International Cosmic Ray Conference,\\
		30 July- 6 August, 2015\\
		The Hague, The Netherlands}

\begin{document}

\section{Introduction}\label{intro}
In cosmic ray (CR) astrophysics, the word ``anomaly'' is a sort of {\it catchall}, used in different contexts and by different people with different meaning. In particular, it  has also a proper,
technical meaning. This is why an initial disclaimer is mandatory: despite the title, in the following I will not discuss about ``Anomalous Cosmic Rays'', or ACR. This is  a peculiar  component (in terms of composition) of low-energy CRs, interpreted as originating from neutral atoms in the interstellar space leaking into heliosphere, that then get ionized, picked up by solar wind and accelerated for instance at the termination shock, eventually propagating back in the inner heliosphere via diffusion and drift (for a review see e.g.~\cite{K99}).

Actually, when I was invited to talk about recent CR anomalies and their interpretation,  this possible confusion due to the by now conventional nomenclature in low-energy CR astrophysics stimulated me to reflect on the meaning to assign to the word {\it anomaly}. By definition, it refers to something unusual and unexpected with respect to the common understanding of a subject. But what ``standard expectations'' are can be quite subjective, assuming a different meaning if talking to experimentalists or theorists, experts of the interstellar medium and propagation or acceleration ones, people focusing on low-energies or high energies, etc.

To narrow down the subject, I decided to limit myself to introducing the recent observations which seem to defy expectations about (Galactic) CRs, focusing on the direct detection energy range (rather than results of indirect techniques studying extensive air showers, particularly at ultra-high energies) and on charged particles only, leaving out neutrinos and photons. In particular, I will indulge in theoretical ideas and models that have been stimulated in the past few years, notably after the presentation of intriguing results at the last couple of ICRC meetings.

\begin{figure}[!tb] 
\centering
\includegraphics[width=0.7\textwidth]{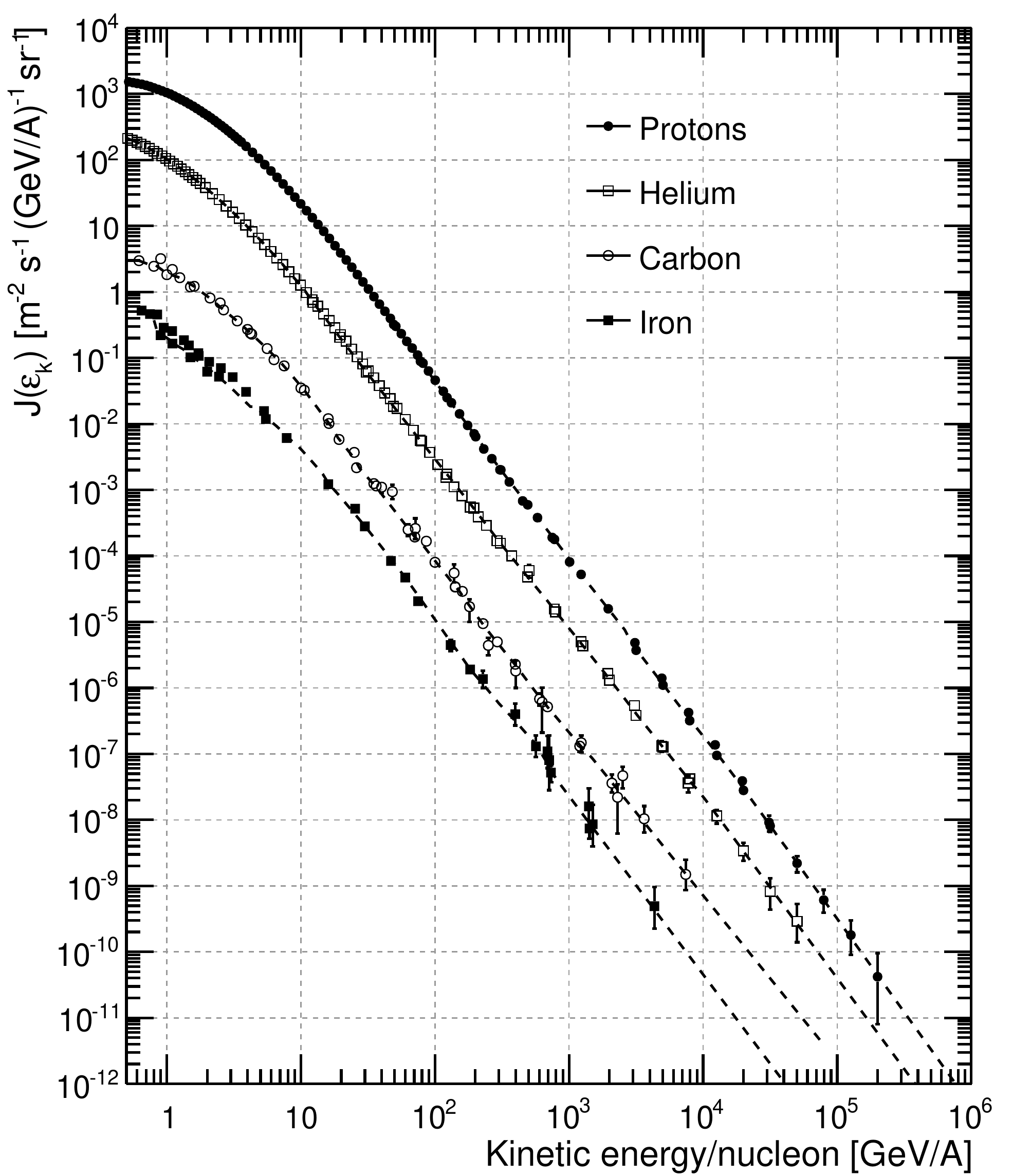} 
\caption{Spectra of some of  the most abundant  cosmic-ray species, from~\cite{Baldini:2014zya}.}
   \label{fig:1}
\end{figure}

To start with the largest possible consensus about what to declare ``an anomaly'' in CR physics, I will recall the basic  {\it common knowledge} about CRs---the 0$^{\rm th}$ order picture taught in introductory university courses---that, starting from ${\cal O}$(10) GeV/nuc, CRs have {\it featureless and universal power-law energy spectra}. This is nicely illustrated in Fig.~\ref{fig:1}, taken from the recent pedagogic review~\cite{Baldini:2014zya} based on the database~\cite{Maurin:2013lwa}.
In fact, power-laws are ubiquitous in Nature as a manifestation of scale-free phenomena. They have been a guiding principle for a number of fruitful theoretical ideas related to CR physics, like  the Fermi acceleration mechanism (where a power-law spectrum follows from the energy-independence of the relative gain of energy per acceleration cycle), or the Kolmogorov theory of turbulence, a benchmark model for diffusive propagation in the interstellar medium (ISM). 
 As in other fields of physics, however, we hope that the identification of departures from simple and universal behaviours might offer the possibility to learn about specific scales and processes peculiar of CR phenomena, shedding light on non-universal features of injection, acceleration, escape, propagation. 

To illustrate this point, I would like to to draw an analogy based on well known physics, which also gives me the opportunity for a homage to a remarkable physicist related to our host country and city. By mid-nineteen century, thanks to the work of Clapeyron and others, it had been recognized that gases---at least for common conditions of (relatively high) temperature and (relatively low) pressure---behave thermodynamically in a universal manner, encoded in the equation of state for perfect  gases, $p\,V=n\,R\,T$,
linking pressure, volume, temperature and molar content without any  gas-specific parameter. 
A few decades later, in 1873, Van der Waals presented his seminal dissertation~\footnote{Most of the work entering his thesis was in fact performed while he was a physics teacher in The Hague.} including among other things the the correction to this law,
\begin{equation}
\left(p+\frac{a\,n^2}{V^2}\right)(V-n\,b)=n\,R\,T
\end{equation}
which applied to a much greater range of conditions, and explicitly included gas-specific parameters, $a$ and $b$, linked to the pressure contribution due to  inter-molecular forces and the finite volume of the molecules, respectively. Besides giving a tool to better reproduce experimental data, however, in his 1910 Nobel Lecture~\cite{Nobel1910} he
clearly  identified his crucial conceptual contribution:
{\quote \it It does not seem to me superfluous, perhaps it is even necessary, to make a general observation [\ldots] in all my studies I was quite convinced of the real existence of molecules, that I never regarded them as a figment of my imagination [\ldots] When I began my studies I had the feeling that I was almost alone in holding that view [\ldots] now I do not think it any exaggeration to state that the real existence of molecules is universally assumed by physicists. Many of those who opposed it most have ultimately been won over, and my theory may have been a contributory factor.  And precisely this, I feel, is a step forward.}

We can only hope  to emulate his sharp vision, trying to infer from anomalies identified in CR data not merely ``better fitting formulae'', but a deeper understanding of the physics processes ruling Galactic CR phenomenology.

\section{The observational situation}\label{observ}
In recent years, the energy range around the TeV/nuc has begun to be explored with sufficient precision that hints of possible departures from extrapolations of lower energies spectra have emerged. 
In particular, this is  the case of the data reported by ATIC-2~\cite{Atic09}
and, especially, by  CREAM~\cite{Cream11}. Power-law fits to data below $\sim 100\,$GeV/nuc (from AMS-01, BESS, Caprice\ldots), when extrapolated to the TeV/nuc region, appear clearly to undershoot the data. The other intriguing fact is that this conclusion seems to apply universally to $p$, He, as well as heavier nuclei! At the same time, the data also suggest that the proton spectrum is softer (by roughly 0.1 in spectral index) than the He one~\cite{Ahn:2010gv}. 
However, it is fair to say that different experiments covering the same energy range still presented a dispersion of results comparable to the size of the discrepancy. This type of considerations (and some familiarity with the challenges found in CR measurements) 
certainly raised the doubt that these potentially interesting observations were not robust with respect to underestimated experimental systematics. After all, experiments do differ among themselves in the technologies used and, sometimes, even in the medium in which they operate (atmospheric balloons vs. space experiments, for instance).

This is why that results presented by PAMELA in 2011~\cite{Adriani:2011cu} were considered an important breakthrough: for the first time a single experiment bridging the gap between ``low-energy'' ones and ``high-energy'' ones showed evidence for a broken power-law at the transition rigidity  of few hundreds GV for both $p$ and He. If one thinks of the amount of theoretical speculations on the CR source and propagation diagnostics driven by two other established spectral breaks at higher energies, the ``knee'' and the ``ankle'' (not to mention the GZK cutoff), the interest in the theoretical community for this result needs no justification!

However, as in a good  thriller, AMS-02 {\it preliminary} results presented at the last ICRC in 2013~\cite{AMS2013} made the situation more confused: although the $p$ vs. He spectral difference seemed to be confirmed, no evidence for breaks in the $p$ and He spectrum was apparent. Fortunately, nowadays a happy ending seems to be in view: the recent AMS-02 {\it publication}~\cite{Aguilar:2015ooa} reports  a change of slope in the proton spectrum qualitatively similar to (and to some extent also quantitatively  consistent with) what published by PAMELA, quantified in a hardening of about 0.13 in the spectral index above 300 GeV.
Even for He, the recent status update on the He flux analysis~\cite{Haino15} gives at least a qualitative confirmation (albeit with some quantitative differences) of the break in the He spectrum, too.
It is also reassuring that BESS results recently appeared~\cite{Abe:2015mga} are in good agreement with PAMELA and (still preliminary) AMS ones on the ratio $p$/He, whose power-law decline with energy seems thus uncontroversial.
Although it is perhaps too early to draw strong conclusions, it is fair to say that current observational results seem to confirm the reality of these phenomena: a break in the power-law nature of the observed Galactic CR spectra and the violation of universality of their spectral indices. Both phenomena need obviously to be tackled by CR theorists!

\section{Theoretical interpretations for the broken spectra}\label{break} 
\subsection{Propagation}\label{propbrok}
A first class of solutions attributes the effect to features in the diffusion coefficient, $K$, while retaining for instance simple momentum power-law injection spectra at the sources.
The most appealing aspect of this class of solutions is that it naturally accounts for its universality in rigidity, compatibly with currently available data. The reason of this universality is evident if---neglecting all losses---we write the equations of motion of a nucleus
with mass number $A$  ($m_p$ being the proton mass) and charge $Z$ (in units of the positron charge $e$) in presence of electromagnetic fields in terms of rigidity ${\bf {\cal R}}={\bf p}\,c/Z\,e$
\begin{equation}
\frac{1}{c}\frac{d {\bf{\cal R}}}{dt}={\bf E}({\bf r},t)+\frac{{\cal R}\times {\bf B}({\bf r},t)}{\sqrt{|{\cal R}|^2+{\cal R}_0^2}}\,,\label{rigidityEOM}
\end{equation}
where
\begin{equation}
\frac{1}{c}\frac{d {\bf r}}{dt}=\frac{\bf{\cal R}}{\sqrt{|{\cal R}|^2+{\cal R}_0^2}}\,,
\end{equation}
and we defined a species-dependent parameter
\begin{equation}
{\cal R}_0=\frac{A\,m_p\,c^2}{Z\,e}\,.
\end{equation}
It is clear from Eq.~(\ref{rigidityEOM}) that, in the relativistic limit where the breaks have been detected,
the diffusive propagation (which at fundamental level is purely scattering onto electromagnetic ``small-scale'' fields) does not depend on the parameter ${\cal R}_0$, and thus on the species.

Different models however differ on the origin of the feature in the diffusion coefficient invoked to explain the data. For instance, in~\cite{Tomassetti:2012ga}   it was argued that by assuming that the diffusion coefficient is a non-factorizable function of rigidity and space, it is possible to obtain the desired effect.
In particular, a different rigidity dependence is assumed to take place in the inner halo (close to the Galactic disk) or the outer diffusive halo. These expectations are physically plausible, for example since in the inner zone supernova remnant (SNR) driven turbulence should dominate, while the one associated with CRs is important far away. Such a type of behaviour has in fact been invoked in the past, see for example~\cite{Erlykin:2002qg}, although the model in~\cite{Tomassetti:2012ga} is purely phenomenological and none of the model parameters is predicted a priori.

\begin{figure}[!tb] 
\centering
\includegraphics[width=0.8\textwidth]{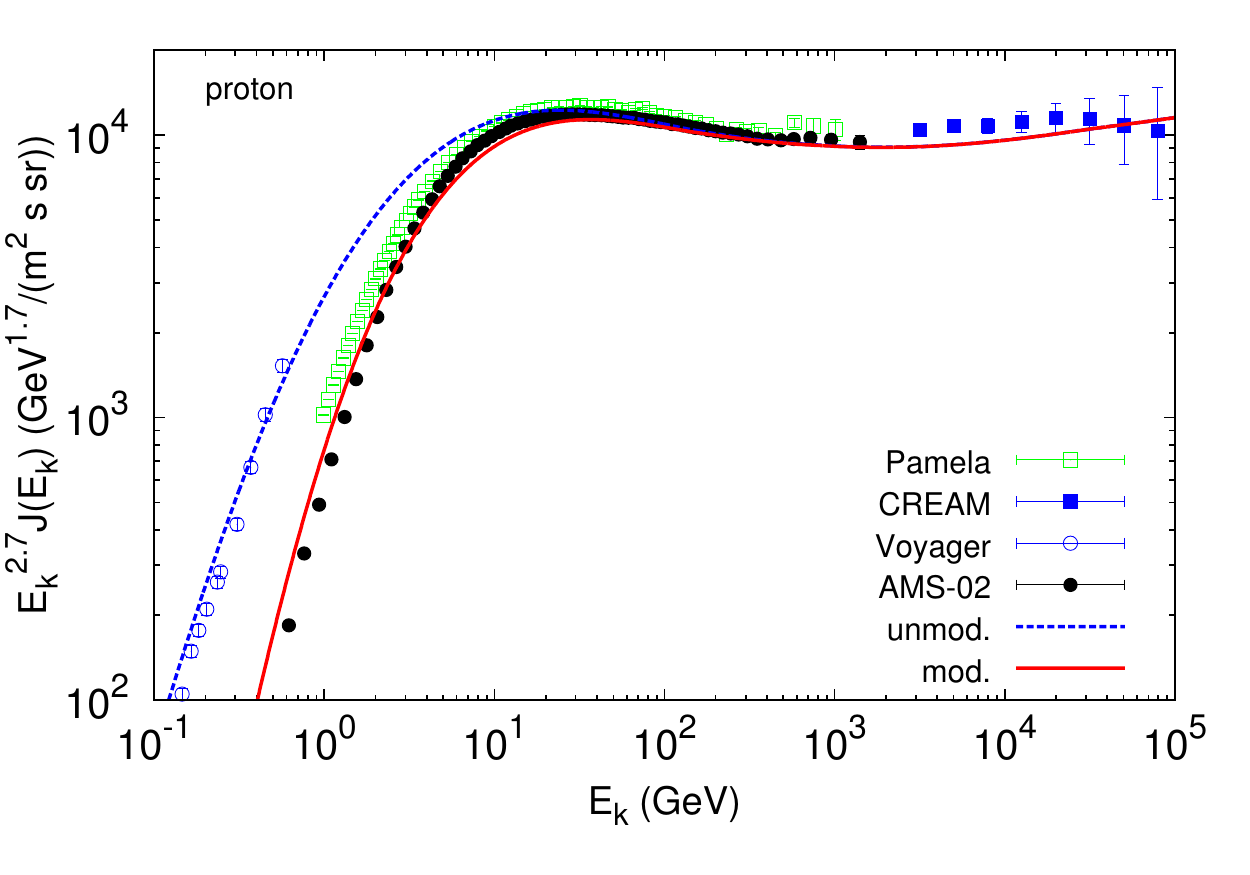} 
\caption{Spectrum  of  protons  measured  by  Voyager 1~\cite{Stone:2013},  AMS-02~\cite{Aguilar:2015ooa},  PAMELA~\cite{Adriani:2011cu} and CREAM~\cite{Cream11},  compared  with the  prediction  of  the  
calculations in~\cite{Aloisio:2015rsa}. The solid line is the flux at the Earth after the correction due to
solar modulation, while the dashed line is the spectrum in the interstellar medium (ISM). Figure from~\cite{Aloisio:2015rsa}.}
   \label{fig:2}
\end{figure}

An alternative model of this class is the one introduced in~\cite{Blasi:2012yr}, which describes the
CRs above the break as diffusing onto external turbulence (as traditionally argued to be injected by SNRs), 
while the change of diffusion coefficient below the break would be due to diffusion onto waves generated by CRs themselves via streaming instability. The break would thus be due to an intrinsically non-linear phenomenon reflecting an improved microphysics description of the CR propagation problem. 
A simple estimate shows that the non-linear Landau damping (controlling the wave diffusion in $k-$space) equals the CR instability growth rate at a rigidity of 200-300 GV, for fiducial values of Galactic parameters, tantalizingly close to the rigidity of the break~\cite{Blasi:2012yr}.
A more complete phenomenological treatment has been performed in~\cite{Aloisio:2013tda} and shown to be in satisfactory agreement with the data available a couple of years ago. The recent publication of
AMS-02 protons~\cite{Aguilar:2015ooa} as well as the important measurements by Voyager 1 of the interstellar medium spectrum of CRs~\cite{Stone:2013}, unaffected by solar modulation, prompted a new look at this model~\cite{Aloisio:2015rsa}, which seems to be in remarkable agreement 
with data of primary CRs over six decades in energy, despite the low number of free parameters (See Fig.~\ref{fig:2}). The agreement with Voyager 1 data is particularly impressive since it comes out automatically and unavoidably, and appears to indicate the correctness of an essentially advective transport (by the self-generated waves) at low energy, as opposed to a transport dominated by large reacceleration effects sometimes invoked in the literature. A dominantly advective transport is also more plausible on the basis of energetics requirements, see e.g. the discussion in~\cite{Drury15} at this conference.

\subsection{Acceleration}\label{accelbrok}
It is logically possible that the observed features are instead imprinted onto spectra {\it at production}. Again, multiple possibilities arise quite naturally. 

For instance, in diffusive shock acceleration (DSA), non-linear effects due to CR feedback are known to
lead to a concavity in the spectrum, which has led some authors to speculate about the relevance
for the problem under discussion~\cite{Ptuskin:2012qu}. In~\cite{Ptuskin:2012qu}, such a solution was found to be viable, although one may have some concerns about the robustness of the feature with respect to a sum over different sources (if anything, due to different SNR ages) and to the feature being reflected also in {\it escaping} particles, not merely the accelerated ones. 

Similar concerns (or, if it turns out to be the solution, similar constraints) are present for another natural solution, relying on the fact that the low energy regime of the particle spectrum may be contributed mostly by relatively old shocks, while young ones should be involved mostly in accelerating high energy particles. Since in standard DSA the injection index $\alpha$ is linked to the Mach number ${\cal M}$ by
\begin{equation}
\alpha= 2\frac{{\cal M}^2+1}{{\cal M}^2-1}\,,
\end{equation}
the natural evolution of a SNR shock from high ${\cal M}$ at early stages to lower ${\cal M}$ at later stages could produce a hardening. Despite being mentioned in several articles (see e.g.~\cite{Ohira:2015ega}) and the appeal of such a general consideration, no quantitative realistic modeling of this scenario seems to be present in the literature. Possibly, it turns out to be hard to model a clear spectral break in this way: a rather smooth change of slope extending over several decades of energy is probably easier to obtain.

A completely different direction in acceleration-related models is the possibility that the change of slope is not reflecting a generic property, but comes from the superposition of populations of particles accelerated at different sources (or different sites of the same source). This may find a concrete astrophysical realization in the environments of OB associations, Superbubbles, or  Wolf-Rayet stars, as discussed in the past, e.g.~\cite{Stanev:1993tx}  or~\cite{Parizot:2004em}, and has been recently revived in
the context of spectral CR anomalies in~\cite{Ohira:2015ega}. 

Yet another possibility~\cite{Thoudam:2014sta} is that the spectral shape arises in the context of a ``weak'' reacceleration (in the sense of ref.~\cite{seo1994}),  associated to the volume of ISM occupied by SNR shocks (mostly old,
low ${\cal M}$), although it has been argued in~\cite{Erlykin:2015mea} that the parameters required may be too extreme to be compatible with the relatively hard $p$ and He spectra.

\subsection{Local Sources}
A further class of models, to the opposite limit of generic mechanisms previously discussed, invokes a prominent effect of local sources. This has been given a remarkable attention in the past few years, see for instance~\cite{Blasi:2011fi,Blasi:2011fm,Thoudam:2011aa,Bernard:2012wt,Bernard:2012pia,Thoudam:2013sia,Liu:2014hra}, and the results can be summarized as follows: relatively rare fluctuations may lead to spectra consistent with observations (with younger nearby objects accounting for the harder spectrum at high energy), although typically at the expense of requiring a rather steep dependence of the diffusion coefficient from the rigidity, say with power-law index $\delta \simeq 0.6$, and/or relatively low frequency for the SN explosions ($\sim$1 per century), and/or a relatively thin diffusive halo. Typically these are in tension with other observations, for instance being associated to a too large predicted anisotropy~\cite{Blasi:2011fm}. At more fundamental level,  at the highest energies the self-consistency of these scenarios is very shaky, as noted in~\cite{Blasi:2011fi}, since
the mean-free-path length in the diffusive motion can be estimated as
\begin{equation}
\lambda\sim 1\,{\rm pc}\left(\frac{H}{\rm 3\,kpc}\right)\left(\frac{\cal R}{\rm 3\,GV}\right)^\delta\,.
\end{equation}
Already at ${\cal R}={\rm 300\,GV}$, where the break is located, one deduces $\lambda\gtrsim 10\,$pc, thus preventing a reliable  diffusive description of sources located within a distance $d\lesssim 100\,$pc, since the diffusive approximation condition $d\gg \lambda$ would not be fulfilled. Especially predictions for the multi-TeV energy range covered by experiments like CREAM and invoking sources as close as Vela, at $d\sim 200\,$pc, appear unreliable.

A different perspective has been taken in~\cite{Tomassetti:2015cva}, where in order to revive the possibility for a hadronic origin for the positron hardening,
a {\it local and old} source contribution is invoked, whereas young, distant SNR dominate primary fluxes at high-energy. The low-energy component in the $p$
and He spectrum would be thus a local phenomenon, connected however with the CR lepton ``anomalies''. Nuclei and ${\bar p}$'s would not show
a corresponding rise since dominated by the far away {\it younger} sources. While phenomenologically viable, this scenario unfortunately loses one clear
prediction of the hadronic models for the CR positron rise (corresponding features in hadronic secondaries) and makes it harder to test. Also, it remains to be
estimated how likely such a superposition happens to be, starting from realistic time and space distribution of the CR sources.

\section{Theoretical interpretations for non-universal spectral indices}\label{nonuniv} 
For the same reasons  discussed in Sec.~\ref{propbrok}, non-universal spectra cannot be attributed to
electromagnetic effects intervening when the particles are already relativistic.  
Therefore, perhaps not surprisingly, a first possibility envisaged to account for the observations was to invoke non-electromagnetic effects in propagation, notably losses via 
spallation~\cite{Blasi:2011fi}. However, the required parameters appeared immediately in tension with secondary nuclei observations, in particular the Boron over Carbon ratio, B/C,
and this possibility has been soon excluded~\cite{Vladimirov:2011rn}. 

An obvious way out would be to invoke different sources or sites for the different species, as for instance in ref.~\cite{Zatsepin:2006ci}. Note however that the compatibility with
observations  imposes some special conditions, notably significantly different $p$/He acceleration efficiency at the two sites, despite these elements
being omnipresent in the ISM. Also, most likely this class of solutions requires the spectral breaks to be propagation-induced, to  
explain their common rigidity (at least for $p$ and He).
  
Perhaps the most appealing solution would be to exploit the ``natural'' evolution of Mach number within DSA already mentioned in Sec.~\ref{accelbrok}.
For some reason, He would be mostly accelerated ``early on'' (when ${\cal M} \gg$1) while $p$'s would have a significant component accelerated
when  ${\cal M}$ is relatively low. Different models of this class differ on the specific reason for this difference:
\begin{itemize}
\item {\it efficiency of injection in the acceleration cycle.}\\
In~\cite{Malkov:2011gb}, it was argued that the Alfv\'en waves at the shock are sourced by the more abundant $p$'s; for the same velocity, injected He ions have twice the gyroradius
and are more likely to return upstream, thus being more efficiently accelerated. 
Although both $p$ and He efficiencies decline with ${\cal M}$, it is argued in~\cite{Malkov:2011gb} that the decline is expected to be steeper for $p$'s.
The combined effect would result into harder spectra for He than for $p$. While conceptually appealing, there is no ``standard theory'' for the injection efficiency
as a function of environmental parameters and even in~\cite{Malkov:2011gb} the authors  had admittedly to rely on a model. Unfortunately, the microphysics of the problem does
include  also a number of other complications, such as the role of partially ionized atoms (see for instance~\cite{Morlino:2010am}), 
and the role of acceleration from dust grains as opposed to gaseous material, see e.g.~\cite{Ellison:1997vx}. Probably, significant theoretical
progress is needed before reassessing the viability of this interesting proposal. 
\item {\it Variable He/$p$ ion concentration in the medium swept by shocks.}\\
One cause of this difference may be a ``time-dependent'' ionization state. It has been argued in~\cite{Drury:2010am} that it is 
expected that older (thus weaker) shocks propagate in a medium where more He is neutral than in the strongly ionized environment of young remnants.
Alternatively, this may ``just'' reflect the chemical environment of the main sources of CRs:
This has been recently discussed in~\cite{Ohira:2015ega} to support a CR origin in superbubbles, following the calculation already presented in~\cite{OI11}.
In this scenario, the authors argue that the abundance of He is naturally expected to be outward-decreasing with respect to the initial shock location. Note however that the presence of 
``spatial segregation'' of He vs. $p$, with the former ions preferentially attained earlier on by the shock, has been argued in~\cite{Erlykin:2015mea}
to be a rather generic expectation, notably in remnants originating from type II supernovae. 
\end{itemize}

\section{Towards testing the models}\label{tests}
Is it possible to conclusively test (some of) the models discussed in Sec.s~\ref{break},\ref{nonuniv}?

At least for the case of spectral breaks, it is quite consensual that the main diagnostics potential comes 
 from secondaries, notably (but not exclusively!)  B/C. In short
\begin{itemize}
\item A source origin for the break implies that the spectra of secondaries directly map the break present in their primary parents spectra. Most secondaries
originate from spallation reactions, which to a good approximation preserve the energy per nucleon, $E/A$. Since this quantity {\it in the relativistic regime} 
is proportional to ${\cal R} \times Z/A\simeq {\cal R}/2$, the conservation of the energy per nucleon also leads to similar mapping of rigidity features. 
Put otherwise, rigidity spectra of most secondaries/primaries should be almost featureless. 
\item If the break originates from propagation effects,  on the top of the feature inherited from parent nuclei, 
secondary nuclei get a further hardening due to subsequent propagation in the ISM.  As in conventional wisdom, at high energies ratios like B/C  reflect 
the rigidity dependence of the diffusion coefficient. In this class of models, secondary spectra should thus show a more pronounced break than primary ones.
\item Models where the break in primary spectra do not reflect their average ISM spectrum, rather a local fluctuation, by construction do not
have a unique, deterministic prediction for the observed flux of secondaries. Nonetheless, in the limit where secondaries are generated by the 
spatially averaged, featureless spectrum of primaries, a {\it softening} is expected in the secondary/primary ratios, since secondary spectra would be pure
power-laws while primaries show a hardening.
\end{itemize}
At present, published data are not precise enough to allow for clear conclusions to be reached. In fact, forecast studies presented at this conference by S. Kunz have shown that 
one needs ${\cal O}$(10\%) precision at $\sim$1TeV/nuc for this purpose~\cite{Kunz15}. Certainly, the quite prominent break shown by preliminary results
on the lithium spectrum presented by AMS-02~\cite{Derome15} does not seem to favour the local source scenario, and would qualitatively point towards
the propagation one. On the other hand, no clear break is visible in preliminary data on the B/C~\cite{Oliva15} which would be compatible with a 
source origin as well, or perhaps even better than with a propagation hypothesis. Definitely, current uncertainties are too big to draw strong conclusions,
although there are realistic possibilities that a conclusive argument in favour of a {\it class} of scenarios for the break may be eventually reached 
with the statistics collected by AMS-02 in its lifetime.

\begin{figure}[!tb] 
\centering
\includegraphics[width=0.8\textwidth]{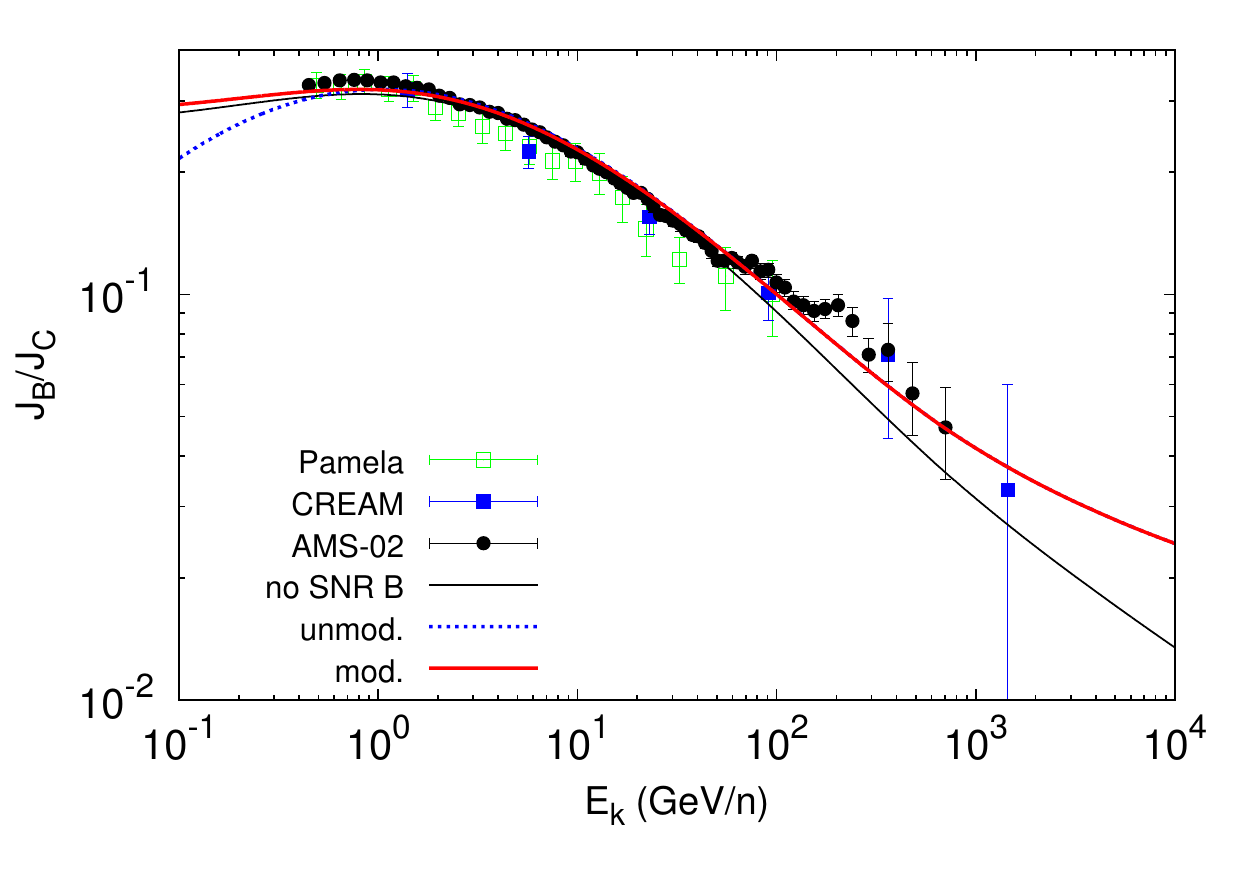}
   \caption{B/C ratio  as measured by CREAM~\cite{Ahn:2008my}, PAMELA~\cite{Adriani:2014xoa}, and according to preliminary measurements of AMS-02~\cite{Oliva15}, compared with the the predictions of the model~\cite{Aloisio:2015rsa}  with (red curve, top) or without (black curve, bottom) adding  a source grammage of $0.15\,$g\,cm$^{-2}$. From~\cite{Aloisio:2015rsa}.}
              \label{fig:3}%
    \end{figure}

On the other hand, distinguishing among models within the same class appears challenging within foreseeable sensitivity, since to large extent their
signatures are degenerate. The hope remains to reduce such degeneracies in models which predict features in associated observables ({\it multimessenger} approach).
Another caveat one should be aware of concerns the role of ``secondary'' production at sources. At energies above $\sim $100 GeV/nuc, it has been shown in~\cite{Aloisio:2015rsa} 
that the grammage expected in typical SNRs, larger than $\sim$ 0.1 g/cm$^2$, is comparable or larger than the size of preliminary error on the B/C data (see Fig.~\ref{fig:3}). Accounting for this contribution may be crucial to assess the viability of the models when comparing  theory with observations.
Actually, this is a generic caveat on the intrinsic limitations in extracting propagation parameters from secondary/primary ratios, notably B/C, as analyzed in~\cite{Genolini:2015cta}
and discussed at this conference.

The situation for the non-universal spectral indices is, if possible, more delicate. Virtually in all scenarios, the indices of the spectrum of the {\it nuclei} (thus including heavier primaries!)
provide essential information for diagnostics. Is the He index identical to the C, O\ldots Fe one? Is there a dependence on the volatility or refractory nature of the nuclei, pointing
to some features associated to injection or the acceleration environment?
While forthcoming data (starting from AMS ones) should help addressing these questions, it is unclear at present if the precision of the experimental determinations on the one hand,
and the accuracy of theoretical calculations, on the other hand, will be sufficient to disentangle among the many scenarios proposed. A definite answer may thus have to wait for many years.

\section{Cosmic ray antimatter}
\subsection{Positrons}\label{posit}
Since at least two ICRC conferences (notably after the first high-significance results reported by PAMELA~\cite{Adriani:2008zr}), the positron fraction has been claimed to show evidence for
an ``anomaly'', commonly known as {\it positron excess}. This is in my opinion a paradigmatic example of a misnomer, since the intriguing observation
is not that the number of positrons (even worse, at the beginning only the positron {\it fraction} had been reported!) is too large compared to what can be produced in hadronic primary collisions with ISM,  but that  their energy spectrum 
is so hard to  require an explanation~\cite{Serpico:2008te}.

The major recent results in this field since the first publication by AMS-02 on the positron fraction~\cite{Aguilar:2013qda} (presented at the last ICRC) have been the data published by AMS-02 on the total flux  of ($e^+ + e^-$)~\cite{Aguilar:2014fea}, on separate $e^+$ and $e^-$ fluxes~\cite{Aguilar:2014mma}, as well the update on the $e^+$ fraction~\cite{Accardo:2014lma}.

From the point of view of interpretations, the situation has only moderately evolved since my review~\cite{Serpico:2011wg} on this topic.
Basically, it seems by now established that the intrinsic positron spectrum is hard up to several hundred GeV at least, comparable in slope with the proton spectrum and thus inconsistent 
with a secondary origin where positrons should have undergone significant steepening especially due  to energy losses (and, to a minor extent, to the rigidity-dependent diffusion). Additionally, the steep spectrum measured in the (dominantly primary) $e^-$ seems consistent with the softening with respect to hadronic CR, due to significant $E$-losses.
As far as I know, there is not anymore a  viable, coherent model of the CR fluxes (i.e. compatible at the same time B/C, the diffuse Galactic gamma-ray spectrum, etc.) explaining what is observed without invoking some primary source of $e^+$, with a relatively hard injection spectrum.
Qualitatively, the properties of these primary $e^+$ sources  are also emerging with more and more clarity, confirming the picture present a few year ago~\cite{Serpico:2011wg}:
\begin{itemize}
\item Positrons must be originating mostly from astrophysical sources rather than from dark matter (DM), notably to fulfill a number of multi-messenger constraints, especially (but not only) from $\gamma$-ray data. In this respect, recent cosmological constraints from Planck have been instrumental in excluding some quite elusive DM models~\cite{Ade:2015xua}.
\item Pulsar wind nebulae are natural astrophysical candidates, since they host a source of relativistic pairs (the pulsar) and are characterized by further acceleration far away from the 
neutron star, notably at the termination shock. All existing information (energetics, spectra) seems compatible with what needed to explain the data, although some important details are still unknown~\cite{Blasi:2010de}. That said, other sources (like SNRs) are certainly possible, at least as sub-leading components (for a review see~\cite{Serpico:2011wg}), and new candidates 
are still being proposed. Even at this conference, several models~\cite{Grimani:2015,Venter:2015} and phenomenological fits~\cite{Boudaud:2015,DiMauro15} have been discussed.
\item Local, discrete sources should be more and more important with energy, and are (at least) naively associated with a ${\cal O}$(0.1\%) anisotropy at ${\cal O}$(100) GeV which might be detectable. 
\end{itemize}
It is worth noting that several of these aspects associated to leptonic spectra were already clear well before  modern data with higher precision became available.
For instance, it was already clear twenty years ago~\cite{Atoyan:1995} that a global understanding of the energy spectra of cosmic ray leptons from sub-GeV to TeV energies
required to account for local (as well as for far-away) Galactic sources, and over a decade ago~\cite{Kobayashi:2003kp} spectral bumps at the TeV energies were already clearly identified as interesting
targets to confirm our expectations on the CR lepton flux. In fact, detecting such features constitutes one of the primary goals of an instrument like
CALET~\cite{CALET15}, which has been recently launched into space and docked to the International Space Station.
Amusingly enough, in the leptonic channel the ``anomaly'' that would require an explanation might be represented by a putative detection of a {\it featureless} power-law extending into the multi-TeV range, exactly the opposite than for the hadronic channel. Needless to say, this peculiarity is not always appreciated!

\subsection{Antiprotons}\label{antip}
Among interesting new results, AMS-02 has presented a {\it preliminary} spectrum of the $\bar{p}/p$ ratio up to $\sim 350\,$GeV~\cite{Kounine15}.
In view of some  old secondary $\bar{p}/p$ prediction band shown together with the results and passing far below the data, it is natural
to ask oneself if the flat behaviour beyond $\sim 10\,$GeV does represent a new intriguing anomaly. 

Even forgetting the preliminary nature of these data (and recent experience should suggest to be careful in over-interpreting them, see 
Sec.~\ref{observ}) it is worth remembering that old predictions cannot be consistently overlapped with the new points for a number of reasons: for instance, they do not take into account the hardening in $p$ and He fluxes at high-$E$ recently detected (a point already made in~\cite{Donato:2010vm}); similarly, new data, analyses and insights on $\bar{p}$ production cross sections, whose implications have been studied just last year~\cite{diMauro:2014zea,Kappl:2014hha}, should be taken into account.
\begin{figure}[!t]
\begin{center}
\includegraphics[width=0.8\textwidth]{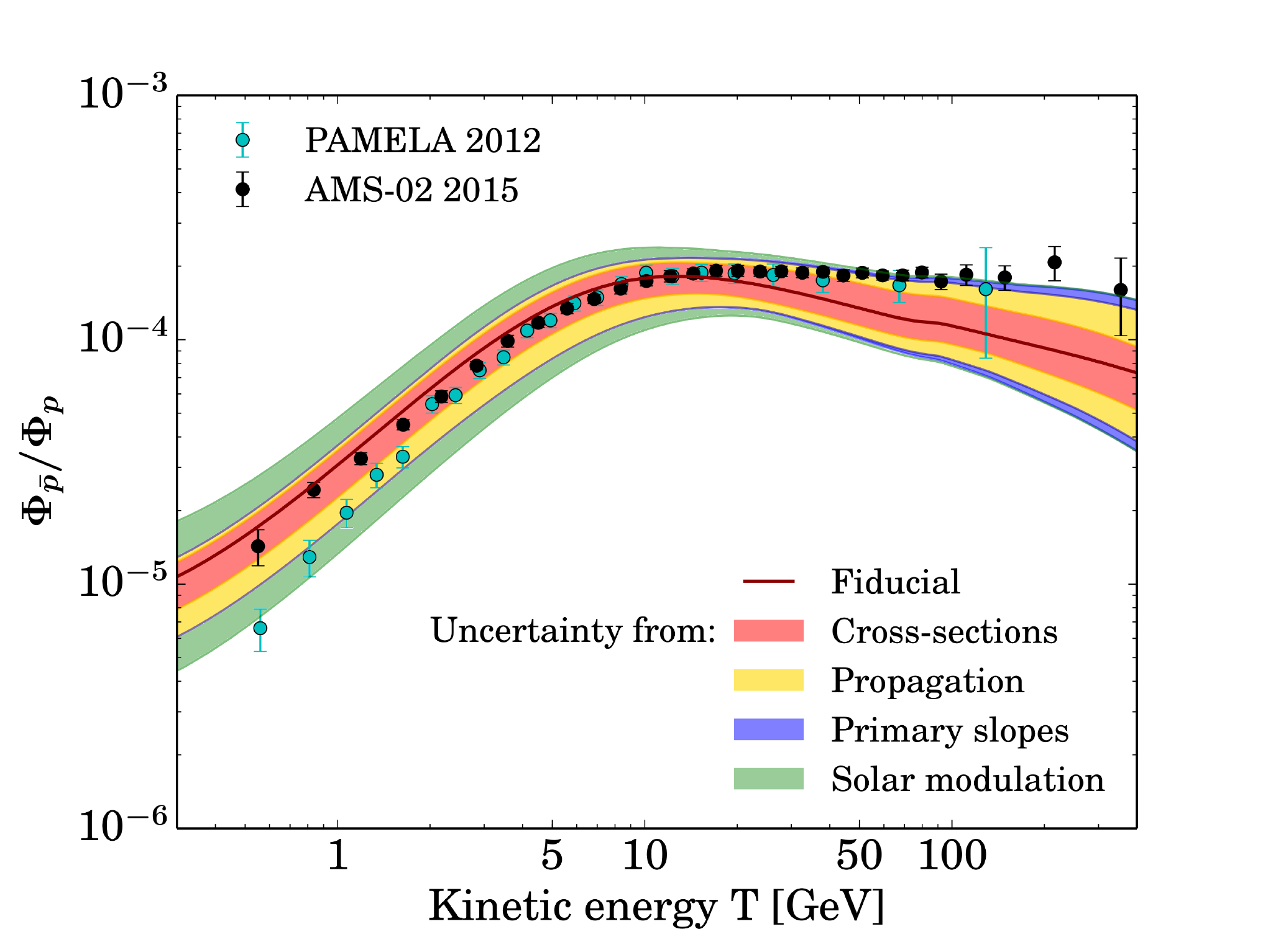}
\caption{\label{fig:4}
The  combined total uncertainty (with different contributions to it) on the predicted secondary $\bar p/p$ ratio superimposed to the older PAMELA data~\cite{Adriani:2012paa} and the new preliminary AMS data~\cite{Kounine15}. From~\cite{Giesen:2015ufa}.}
\end{center}
\end{figure}

In~\cite{Giesen:2015ufa}, the secondary $\bar{p}/p$ background was recalculated {\it within old propagation models}, while updating these physics inputs and with a realistic account of uncertainties. Comparing the new prediction with the data (see Fig.~\ref{fig:4}) one deduces that an excess (or ``antiproton anomaly'') cannot be unambiguously established for the time being. Although most models pass below the data at high energy, some configurations are consistent with them within errors. If confirmed, however, AMS-02 data would definitely prefer propagation models with a mild dependence of the diffusion coefficient on rigidity, probably  $\delta\lesssim 0.4\div 0.5$.
Some other works~\cite{Evoli:2015vaa,Kappl:2015bqa} have attempted to revisit also the propagation model parameters on the light of new results on B/C (the ones published by PAMELA~\cite{Adriani:2014xoa} and the preliminary B/C data presented by AMS-02~\cite{Oliva15}): in both cases they reach similar if not stronger conclusions, hinting at their robustness.

Antiproton data with the precision and the dynamic range within reach of AMS-02 are potentially of great importance both for the validation of CR propagation models and for indirect
searches of DM (a summary of the presentations on this topic at this conference has been provided by M. Cirelli).  While the exploration of the hundreds of GeV range is ushering us into a new era, a caveat similar to the
one raised in Sec.~\ref{tests} applies here as well: just as B/C, also $\bar{p}$'s at high energies could be significantly affected by production at sources. This was already noted in the
past, see for instance~\cite{Blasi:2009bd}, which stressed the important degeneracy for DM searches (later confirmed explicitly in~\cite{Pettorino:2014sua}). 
New calculations of secondary production at sources have been presented at this conference, too~\cite{BK15}.

\section{Discussion and conclusion}
I have reviewed some intriguing ``anomalies'' in the field of charged cosmic ray astrophysics, unveiled by recent direct measurements (Sec.~\ref{observ}).
Earlier contradictory results on the presence of spectral breaks seem now to converge towards a consensus on their reality at rigidities
of $\sim 300\,$GV, at least for protons and He. Spectral differences (at least between $p$ and He) seem also unambiguously established.
I then highlighted a number of hypotheses for the origin of these features (Sec.s~\ref{break},\ref{nonuniv}). The importance of measurements of nuclei, both
primary and secondary ones, for disentangling among different classes of solutions has been reviewed in Sec.~\ref{tests}. The situation in antimatter,
notably positrons (Sec.~\ref{posit}) and antiprotons (Sec.~\ref{antip}), has been summarized, too. The need for primary positron sources has been more and more
robustly established thanks to new AMS-02 results, with the next frontier being the confirmation that spectral features
in the lepton cosmic ray spectra are present in the trans-TeV range, due to prominent local sources. Despite early impressions, preliminary antiproton results
seem still consistent with pure secondary production within (at least some) existing propagation models. Accounting for secondary production at sources in theoretical predictions
becomes more and more important, for meaningful comparison with precision data at hundreds of GeV/nuc.

The discussion developed in this contribution does not exhaust all ``anomalies'' attributed to CRs, of course. 
For instance, basically all considerations till now pertain to locally observed CR fluxes. 
If CR in the Galaxy were homogeneously distributed, no further complication would arise.
But neither the CR source distribution nor diffusive propagation are expected to be homogeneous, and the diffusion coefficient has in fact
a tensorial (generically anisotropic) nature. In principle, the gamma-ray flux morphology (proportional to a line-of-sight convolution of
the CR flux and target density) and the CR anisotropy (dependent on the {\it gradient} of CRs)
are diagnostics tools to infer  {\it non-local} properties of the CR flux, elsewhere in the Galaxy.
Perhaps not surprisingly---since linked to things we know relatively little about!--- both are associated to long-standing ``anomalies'', the
gamma-ray gradient problem and the anisotropy problem(s), whose nature and possible solutions have been deeply investigated.
At this conference, these issues have been touched upon and discussed e.g. in~\cite{EWD15}, \cite{MA15} or Gaggero's highlight talk, to which 
I address for more details.

My overall impression is that, provided that the precision of modern experiments is not illusory (i.e. systematics are not underestimated), CR physics is 
experiencing a phase of ``normal progress'' in experimental science, where with higher precision one  sees cracks emerging in the simplest theoretical
models used for interpreting the old data.
The violations of species universality and power-laws in CR spectra may be such signs. Let me emphasize that this ``healthy'' progress is rare and extraordinary 
for CR physics, for too long affected by the situation sharply summarized in a question and answer  to be found in T. Stanev's  textbook ``High Energy Cosmic Rays'': 
{\it Is progress in the cosmic ray field slow? It certainly looks like that}. Only thanks to concerted and careful efforts of numerous experimental teams (to whom we should be all grateful) this progress
has become possible.  This situation probably requires new benchmarks in theoretical predictions to emerge, with the old ones becoming too simplistic.

In my opinion, the most pressing issue is to understand how many of these {\it cracks} are telling us something generic about CR sources and propagation, and what
aspects are instead {\it accidental}, for instance depending on the specific position and time of the Galaxy we happen to live in (i.e. distance to the nearest source): while both types of discoveries
would be beneficial, only the former ones are likely to lead to major advances in our understanding of CR-related phenomena, I believe. In this sense, {\it Van der Waals' lesson} recalled in Sec.~\ref{intro} is quite actual and pertinent to our field!

\begin{acknowledgments}
I would like to thank the ICRC 2015 organizers for their kind invitation to talk about this stimulating topic, as well as all my collaborators on the subjects I have touched upon in my talk,  R. Aloisio, E. Amato, M. Boudaud, M.~Cirelli, M. di Mauro, F. Donato, Y. Genolini, G. Giesen, A. Goudelis, V. Poulin, A. Putze, P. Salati, and in particular, P. Blasi. A special thank to the Mainz Institute for Theoretical Physics (MITP) for its hospitality and support during the initial stage of the redaction of this proceeding. Partial support by by the French ANR, Project DMAstro-LHC,  ANR-12-BS05-0006 is also acknowledged. 

\end{acknowledgments}

\end{document}